\documentclass[onecolumn,draftclsnofoot, 12pt]{IEEEtran}
\usepackage[T1]{fontenc}
\usepackage[latin9]{inputenc}
\usepackage{cite}
\usepackage{amsmath,setspace}
\usepackage{amssymb}%
\usepackage{graphicx,subfigure}
\usepackage{empheq}
\usepackage{color,upgreek}
\newcommand{\bi}[1]{\boldsymbol{#1}}
\renewcommand{\vec}[1]{\bf{#1}}     %equation styles %vector/matrix
\newcommand{\herm}{^{\mbox{\scriptsize H}}}
\newcommand{\trans}{^{\mbox{\scriptsize T}}}

\begin{document}
\title{Computationally Efficient Robust Beamforming for SINR Balancing in Multicell Downlink}
\author{Muhammad Fainan Hanif, Le-Nam Tran,~\IEEEmembership{Member,~IEEE},
    Antti T\"olli,~\IEEEmembership{Member,~IEEE}, and Markku Juntti,~\IEEEmembership{Senior Member,~IEEE}
  \thanks{The authors are with the Department of Communications Engineering and Centre for Wireless Communications, University of Oulu, Finland. Email: \{mhanif, ltran, atolli, markku.juntti\}@ee.oulu.fi.}
}
\maketitle
\thispagestyle{empty}
\vspace{-40pt}
\begin{abstract}
We address the problem of downlink beamformer design for signal-to-interference-plus-noise ratio
(SINR) balancing in a multiuser multicell environment with
 imperfectly estimated channels at base stations (BSs). We first present a
semidefinite program (SDP) based approximate solution to the
problem. Then, as our main contribution, by exploiting some
properties of the robust counterpart of the optimization problem, we
arrive at a second-order cone program (SOCP) based approximation
of the balancing problem. The advantages of the proposed SOCP-based design are twofold. First, it greatly reduces the computational complexity compared to the SDP-based method. Second, it applies to a wide range of uncertainty models. As a case study,
we investigate the performance of proposed formulations when the base station is equipped with a massive antenna array.
Numerical experiments are carried out to confirm that the proposed robust designs achieve favorable results in scenarios of practical interest.
\end{abstract}
\section{Introduction}
In practical wireless systems, it is virtually impossible to provide an error-free estimate of channel state information (CSI) to the transmitter. Although beamforming is very
attractive from implementation and performance perspective, its applicability is reduced due to its
sensitivity to channel estimation errors which may arise as a consequence of pilot contamination in multicell systems \cite{Jose}, quantization effects due to digital processing \cite{suraweera} etc.
Motivated by this dilemma, various studies have been conducted to design `uncertainty immune'
precoders, see e.g., \cite{vorobyov,boche,gan,tajer} and
references therein. The key tool common to all the studies
is the application of various important results from the robust
optimization theory \cite{bertsimas1,nemirovski2}. For any optimization problem, the design of robust
counterpart can potentially suffer from two major
difficulties, namely, (i) hurdles in obtaining tractable
representation of the robust counterpart of the original program
thereby compelling to employ various approximations, and, (ii)
once a tractable formulation is obtained an increase in the complexity of the robust counterpart is seen as compared to
the original problem. This  pattern is common to most robust designs pertaining to
signal processing and communication applications in the  literature.\par
The significance of (ii) in designing uncertainty immune precoders is further enhanced when
some of the parameters involved in the system setup take very large values. In this context, the recently envisaged large scale massive
multiple-input multiple-output (MIMO) systems \cite{rusek} can be considered. Indeed, such large-scale antenna arrays promise
increased link reliability, better spectral efficiency and low power consumption at the cost of manyfold increase in the
number of transmitter antennas compared with the traditional multiple-antenna systems. For instance, values of the order of hundreds of base station
antennas have been proposed in \cite{rusek}. The sub-optimality of traditional precoding methods like zero-forcing, block-diagonalization is now
well understood \cite{tran}. Any algorithm that is, for example, based on traditional mathematical programming is likely to outperform the heuristic approaches
of zero-forcing etc. It is also pertinent to point out that the
mathematical analysis of the present paper can be easily leveraged to the case of
maximizing weighted sum rates based on the development presented in \cite{tran}. The traditional approaches of introducing robustness in the the precoder design can end up in a semidefinite program (SDP). The complexity of an
SDP is highly sensitive to the precoder size (more details on this appear in Sec.~\ref{Red_Comp}), and hence the SDP-based solutions can either
incur appreciable computational cost or in certain circumstances the digital resources may not be sufficient to cater for the memory requirements
of an SDP-based solution. On the other hand, second-order cone programs (SOCPs)  are much more computationally efficient (again the details appear in Sec.~\ref{Red_Comp}), and
can certainly provide a viable alternative to designing algorithms for very large-scale antenna arrays. This motivates arriving at robust SOCP formulations
for optimizing certain performance metric in modern communication systems.
\par
In this paper, we study the problem of signal-to-interference-plus-noise ratio
(SINR) balancing in multicell multiple-input single-output (MISO) downlink or interfering
broadcast channel with a realistic assumption of imperfect CSI. We focus on centralized base station (BS) control. The
worst case design philosophy that is commonly employed in the existing
literature is considered.  We first show that the
robust counterpart can be relaxed to an SDP,
and, thus, can be (suboptimally) solved in conjunction
with a bisection search. It appears that the SDP-based formulations provide a general solution to the robust design in many works, e.g., \cite{tajer,gan,boche}. However, the SDP-based methods
may rely on a rank relaxation scheme, which is in general a suboptimal technique.
Furthermore, the SDP-based approaches generally result in computationally
expensive tractable robust counterparts.
As our main contribution, we propose a robust design which is merely
based on solving SOCPs, i.e., the proposed method does not represent
much increase in complexity in comparison to the original version of
the SINR balancing problem. This is accomplished by exploiting
various properties of the constraints in the robust counterpart of
the balancing problem. In particular, we avoid formulating the beamformer design by projecting it
to the space of semidefinite matrices which normally results in a rank
constrained SDP. More importantly, the proposed SOCP-based design
can be used in a wide range of uncertainty models.
We notice that the SDP-based design formulations commonly used in literature \cite{gan,tajer,boche} are only applicable to the cases
where the channel errors lie in an ellipsoid. As mentioned above, we also compare and contrast
the SDP and SOCP solutions, particulary from the computational cost perspective, when the number of base station antennas is very large \cite{rusek}. Finally, through
numerical investigations, we show that the proposed SOCP-based
solution offers comparable performance to the approach in \cite{tajer} and the SDP-based
method when same uncertainty set (a ball) is used to represent channel perturbations.\par

The rest of the paper is organized as follows. Section~\ref{pro_for}
presents problem formulation, a solution for perfect CSI, and
modeling of the balancing problem with imperfect CSI.
Section~\ref{imp_CSI} discusses in detail various solutions with
imperfect CSI along with a comparative discussion about their
properties. Finally, Sections~\ref{res} and \ref{con} describe
numerical experiments and conclusions, respectively.\footnote{We use bold lowercase letters to express vectors and bold
uppercase letters to represent matrices. {\color{black}$(.)\herm$,
$(.)\trans$ and $\mathrm{Tr}(.)$ represent the Hermitian, transpose
and the trace operators, respectively}. $\mathbb{C}^{a\times b}$ and
$\mathbb{R}^{p\times q}$ represent the space of complex and real
matrices (vectors) of dimensions given as superscripts,
respectively. $|\mathcal{M}|$ denotes the cardinality of  set
$\mathcal{M}$. {\color{black}$[{\vec p}]_k$ represents the
$k$th component of vector ${\vec p}$. $|c|$ and $\Re(c)$ represent the absolute
value and the real part of a complex number $c$, respectively}. {${\vec I}_T$ denotes a $T\times T$ identity matrix.} Finally, $\|.\|_2$ represents the $l_2$
norm.}

\section{System Model}\label{pro_for}
\begin{figure}
\centering  \includegraphics[clip,width=0.6\columnwidth]{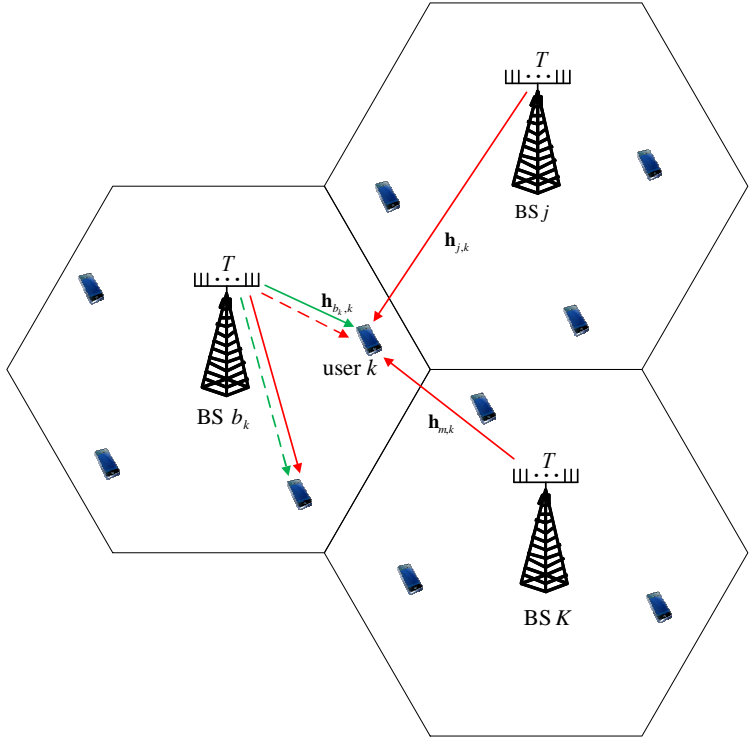}
\caption{System model of a multicell MISO downlink channel. Green lines
represent desired signals, while red ones denote interference.
The serving base station for user $k$ is denoted by $b_{k}$.}
\label{fig:systemmodel}
\end{figure}
Consider a system of $B$ coordinated BSs and $K$ users where each user is served by
one BS. Each BS is
equipped with $T$ transmit antennas and each user with a single
receive antenna. A sketch of the system model is presented in Fig. \ref{fig:systemmodel}. Interference originating outside the coordinated
system is omitted.
The serving BS for the $k$th user is denoted by $b_k$. The signal received by the $k$th user is %given by
\begin{empheq}{align} \label{eq:system_model}
y_k &= {\vec h}_{b_k,k} {\vec x}_{k} +  {\sum\limits_{i=1, i\neq
k}^{K}}  {\vec h}_{b_i,k} {\vec x}_{i} +  n_{k}
\end{empheq}
where ${\vec h}_{b_k,k} \in \mathbb{C}^{1 \times T}$ is the channel (row)
vector from BS $b_k$ to user $k$, ${\vec {x}}_{k} \in
\mathbb{C}^{T\times 1}$ is the transmitted signal vector from the
BS $b_k$ to user $k$ and $ n_{k} \sim \mathcal{C} \mathcal{N} (0,
\sigma^2 )$ represents circularly symmetric zero mean complex Gaussian noise with variance $\sigma^2$.
The transmitted signal vector is defined as ${\vec x}_{k}  =
{\vec{m}}_{k} {d}_{k}$, where ${\vec{m}}_{k} \in \mathbb{C}^{T\times
1}$ is the unnormalized beamforming vector and $d_{k}$ is the
normalized complex data symbol. The total power transmitted by
BS $b$ is $\sum_{k \in \mathcal{U}_{b}}{\mathrm{Tr}}
\left({\mathrm{E}} \left[ {\vec x}_{k} {\vec x}_{k} \herm \right]
\right)=\sum_{k \in \mathcal{U}_{b}}
\bigl\|{\vec{m}}_{k}\bigr\|_2^2$, where the set $\mathcal{U}_b$ with
size $K_b=|\mathcal{U}_b|$ includes the indices of all users served by  BS
$b$. The SINR at user $k$'s receiver is
\begin{equation}\label{eq:SINR_def}{
\gamma_k=  \frac{\bigl|{\vec h}_{b_k,k} {\vec{m}}_{k}  \bigr|^2 } {\displaystyle\sigma^2 + \sum_{i \in \mathcal{U}_{b_k} \setminus k} \bigl| {\vec h}_{b_k,k} {\vec m}_{i} \bigr|^2 + \sum_{b=1, b\neq b_k}^{B} \sum_{i \in \mathcal{U}_{b}} \bigl|{\vec h}_{b,k} {\vec m}_{i} \bigl|^2 }}
\end{equation}
where the interference power in the denominator is divided into intra- and
inter-cell interference components.
\vspace{0pt}
\subsection{Problem Statement and Solution for Perfect CSI}
For the case of perfect CSI, the maximin SINR balancing can be cast as
\begin{equation}\label{eq:opt_problem}
\begin{array} {ll}
\displaystyle\underset{{\vec m}_k:\sum_{k\in \mathcal{U}_b}\|{\vec m}_k\|_2^2\leq
P_b, \forall b}{\operatorname{maximize}}& \displaystyle\min_{k}\alpha_k\gamma_k
\end{array}
\end{equation}
where $\alpha_k$ are positive weighting factors. Using \eqref{eq:SINR_def}, we can equivalently reformulate  \eqref{eq:opt_problem}
as
\begin{equation}\label{eq:ideal_SOCP}
\begin{array} {ll}
\underset{{\vec m}_k, t}{\operatorname{maximize}} & t\\
\operatorname{subject\;to} & \left\|( {\vec
h}_{1,k}{\vec M}_1 \; \cdots \; {\vec h}_{B,k}{\vec M}_B \;
\sigma)\herm\right\|_2\leq\sqrt{1+\frac{\alpha_k}{t}} |{\vec h}_{b_k,k} {\vec m}_{k}|, \; \forall \ k,\\
&\|\mathrm{vec} ({\vec M}_b)\|_2\leq \sqrt{P_b}, \quad\forall b.
\end{array}
\end{equation}
where ${\vec M}_b=[{\vec m}_{\mathcal{U}_b(1)}, \ldots, {\vec m}_{\mathcal{U}_b(|\mathcal{U}_b|)}]$ includes the precoders of all users being served in the
$b$th cell and the operation $\mathrm{vec}(.)$ vectorizes the argument matrix by stacking columns.
Furthermore, we can still find an optimal solution of \eqref{eq:ideal_SOCP} even if
 ${\vec h}_{b_k,k}{\vec m}_{k}$, for all $k$, is forced to be real
\cite{tolli,wiesel,vorobyov}. In this way, the constraints in
\eqref{eq:ideal_SOCP} represent second order cone (SOC) constraints
for fixed $t$. Therefore, the original problem can be solved as a
series of SOCP feasibility problems using bisection search
\cite{boyd,tolli}.\vspace{-0pt}
\subsection{Modeling of Imperfect CSI}
In real systems it is
impossible to achieve perfect transmitter CSI due to several reasons
mentioned in, e.g., \cite{tajer,gan}. Hence, robust designs dealing with channel errors are of  practical importance.
We consider the channel uncertainty model in which the true channel vectors
are of the form
\begin{empheq}{align}\label{eq:CSI_uncertainty}
{\vec h}_{b,k}={\vec {\hat
h}}_{b,k}+\sum_{i=1}^{l_{b,k}}{\boldsymbol \delta }_{b,k}^i[{\vec v}_{b,k}]_i
 ={\vec {\hat h}}_{b,k}+{\vec
v}_{b,k}{\vec A }_{b,k},\quad\forall \ b,k
\end{empheq}
where ${\vec {\hat h}}_{b,k}$ represents the nominal (known) value
of the channels, $l_{b,k}\in\{1,2,\ldots,T\}$, the vectors ${\boldsymbol \delta }_{b,k}^i$ (channel perturbation directions)
form the rows of ${\vec A }_{b,k}\in \mathbb{C}^{l_{b,k}\times T}$ and ${\vec
v}_{b,k}{\vec A }_{b,k}$ gives the error
vector in the downlink channel from BS $b$ to user $k$ \cite{bertsimas1,nemirovski1}. We denote by $\mathcal{S}$ the uncertainty set that includes all channel error row vectors
${\vec v}_{b,k}$. As seen in \eqref{eq:CSI_uncertainty} the above
model assumes that the uncertainty vector affects the data in an
affine manner. This philosophy has been widely used, e.g., in \cite{gan,tajer}
etc. In addition to affecting the
true channels in an affine manner, the error vectors are also constrained to lie in an uncertainty set
$\mathcal{S}_{b,k}$ as
\begin{align}
\mathcal{S}_{b,k}=\{{\vec v}_{b,k}:\|{\vec v}_{b,k}\|\leq \rho_{b,k},\forall b,k\}\label{NBSets}
\end{align}
where $\|\cdot\|$ is an appropriate norm specified by the parameter $\rho_{b,k}$, and is chosen based on how one wishes to
model channel uncertainties. Normally, modeling
channel errors by exploiting their statistical nature is prone to numerous difficulties. To name a few, it requires
information about the statistics of the error vectors which is mostly not available on account of myriad of phenomenon
involved in the channel estimation process. Then even if some information about the statistics of
the error vectors is exploitable, a part from simple linear constraints contaminated with Gaussian errors, it is virtually impossible to
arrive at exact tractable versions of \emph{stochastic constraints}. Motivated by the dilemma, the uncertainty in channels
is modeled by norm-bounded sets \eqref{NBSets}. With such modeling, it does not remain necessary to know
information about the, say, probability law that the uncertainty vectors follow. Further to this, as we will see in the
discussion to follow, norm-bounded uncertainty sets model various real world scenarios very well. One more advantage of such modeling is
that in several cases of interest, the norm-bounded uncertainty models permit either exact tractable formulations or good approximations \cite{nemirovski1,bertsimas2}.
\vspace{-0pt}
\subsection{Worst Case Design Formulation}
We will concentrate on the worst case robust
optimization approach of \cite{nemirovski2}, \cite{nemirovski1} that
has been traditionally used in the existing literature for different
problems \cite{tajer,gan,boche}. The worst case approach amounts to
satisfying the constraints for all possible  channel vectors. Hence, the robust
counterpart of \eqref{eq:opt_problem} is written as
\begin{equation}\label{eq:worst_case_orig}
\begin{array} {ll}
\underset{{\vec m}_k,t}{\operatorname{max.}}& t\\
\operatorname{s.\;t.} & \displaystyle \frac{\alpha_k}{t}\bigl|{\vec
h}_{b_k,k}{\vec m}_k \bigr|^2\geq \sum_{i \in \mathcal{U}_{b_k}
\setminus k} \bigl| {\vec h}_{b_k,k} {\vec m}_{i} \bigr|^2+
\displaystyle\sum_{b=1, b\neq b_k}^{B} \sum_{i \in \mathcal{U}_{b}}
\bigl|{\vec h}_{b,k} {\vec m}_{i} \bigl|^2 + \sigma^2,\:\forall \ k
,\:\: \forall \ \{ {\vec v}_{b,k}{\vec A
}_{b,k}:{\vec v}_{b,k}\in \mathcal{S}_{b,k} \} \\
&\displaystyle \sum_{k\in \mathcal{U}_b}\|{\vec m}_k\|_2^2\leq
P_b, \forall b.
\end{array}
\end{equation}
We note that the formulation in \eqref{eq:worst_case_orig} is intrinsically intractable owing to its semi-infinite nature i.e.,
finite optimization variables and infinite constraints.
\newtheorem{remark}{Remark}
\begin{remark}
It is worth mentioning here that for
the case of receivers equipped with multiple antennas (and hence the possibility of transmitting multiple data streams), an option
could be to employ receiver combining matrix and study the balancing problem on per stream basis. The problem is then
exactly similar to our case for a given receiver processing matrix per user. Here we stress that even in the presence of perfect CSI, joint transmitter-receiver processing matrix design is a difficult nonconvex problem.
\end{remark}
%
%\vspace{-10mm}
\section{Robust Beamformer Designs for Imperfect CSI}\label{imp_CSI}
\subsection{SDP-Based Robust Design}\label{Wor_Cs}
The solution to problem \eqref{eq:worst_case_orig} significantly
depends on the type of the uncertainty set. In the context of SDP design, the most commonly
considered uncertainty model is the one where error vectors ${\vec v}_{b,k}$ are bounded in a ball of radius $\rho_{b,k}$, i.e., $\mathcal{S}_{b,k}$ is expressed as
\begin{align}
\mathcal{S}_{b,k}=\{{\vec v}_{b,k}:\|{\vec v}_{b,k}\|_2\leq \rho_{b,k},\forall b,k\}.\label{s1}
\end{align}
It is clear from \eqref{eq:CSI_uncertainty} and \eqref{s1} that ${\vec h}_{b,k}$ is assumed to lie in an ellipsoid centered at ${\hat{\vec h}}_{b,k}$, which is characterized  by  $\rho_{b,k}$ and ${\vec A }_{b,k}$. This type of channel error model is commonly known as ellipsoidal uncertainty model in the related literature. For practical channel estimation schemes it is known that the channel estimation error follows
Gaussian distribution \cite{Yoo}. Most of the probability content of multi-dimensional Gaussian density is
localized in its certain region. This clearly motivates
modeling the error using an ellipsoid. Further, when vector quantization is used at the receiver, quantization errors can also be approximated by ellipsoids \cite{Zheng:2007}. For this specific uncertainty set, we will show that the robust
counterpart in \eqref{eq:worst_case_orig} can be solved using SDP
approximations. We note that in \eqref{eq:worst_case_orig} the
uncertain part of ${\vec h}_{b_k,k}$ varies in the same set on both
sides of the inequality. It is well known that this renders the
problem intractable \cite{boche,bertsimas2,nemirovski2}. However, we
will see that, after a suitable relaxation, this constraint can be
written in a tractable form. To this end, we define ${\vec
P}_k={\vec m}_k {\vec m}_k \herm$, ${\vec Q}_b=\sum_{k \in
\mathcal{U}_b}{\vec P}_k$ and ${\vec W}_k=\frac{\alpha_k}{t}{\vec
P}_k-\sum_{i \in \mathcal{U}_{b_k} \setminus k} {\vec P}_i$. Then by
using slack variables we observe that the
constraints involving perturbed channels can be cast into tractable linear matrix inequalities
(LMIs) using the so called $\mathcal{S}$-Lemma or
$\mathcal{S}$-Procedure \cite{boydLMI,nemirovski2}.
After some manipulations,
\eqref{eq:worst_case_orig} can be equivalently rewritten as
\begin{subequations}\label{11nm}
\begin{align}
&\underset{{\vec P}_k, t_{b,k},\lambda_{b,k},\tau_k}{\operatorname{maximize}}& & t\notag\\
&\operatorname{subject\;to} &&
\begin{pmatrix}
{\vec A}_{b_k,k}{\vec W}_k {\vec A}_{b_k,k}\herm +\lambda_{b_k,k}
{\vec I} &{\vec A}_{b_k,k}
{\vec W}_k {\vec {\hat h}}_{b_k,k} \herm \\
{\vec {\hat h}}_{b_k,k} {\vec W}_k {\vec A}_{b_k,k}\herm & {\vec {\hat
h}}_{b_k,k} {\vec W}_k {\vec {\hat h}}_{b_k,k} \herm
-\tau_k-\lambda_{b_k,k} \rho_{b_k,k}^2
\end{pmatrix}\succeq 0,\label{11a} \ \forall \ k \\
&&&
\begin{pmatrix}
-{\vec A}_{b,k}{\vec Q}_b {\vec A}_{b,k}\herm +\lambda_{b,k} {\vec
I} & -{\vec A}_{b,k}
{\vec Q}_b {\vec {\hat h}}_{b,k} \herm\\
-{\vec {\hat h}}_{b,k}{\vec Q}_b {\vec A}_{b,k}\herm & -{\vec {\hat
h}}_{b,k}{\vec Q}_b {\vec {\hat h}}_{b,k} \herm+t_{b,k}-
\lambda_{b,k}\rho_{b,k}^2
\end{pmatrix}\succeq 0  \ \forall \ k, b\neq b_k \label{11b}
\\
& & & \sum_{k\in \mathcal{U}_b} \mathrm{trace}({\vec P}_k)\leq P_b,\forall b,
\sum_{b=1, b\neq b_k}^{B} t_{b,k}+\sigma^2\leq
\tau_k, \tau_k\geq 0 \ \forall \ k \label{11c}\\
&&& \lambda_{b,k}\geq0 \ \forall \ b,k, \quad t_{b,k}\geq 0 \
\forall \ k, b\neq b_k,\quad {\vec P}_k \succeq 0,
\mathrm{rank}({\vec P}_k)=1 \ \forall \ k.\label{11d}
\end{align}
\end{subequations}
Therefore, after ignoring the nonconvex rank constraints, bisection search
over $t$ can be used to obtain covariance matrices. However, we
cannot guarantee optimality (see \cite{song}) of the proposed
solution if we obtain the rank of precoding matrices greater than
unity. Similar rank relaxation approach was also adopted in the
recent works, e.g., \cite{gan}. We may need some randomization
procedure \cite{luo} to extract the beamformer ${\vec m}_k$ if
$\mathrm{rank}({\vec P}_k)
> 1$. Nonetheless, randomization trick may not always be useful \cite{gan}.
%\vspace{-15pt}
\subsection{SOCP-Based Robust Scheme}\label{Red_Comp}
The robust counterpart of any optimization problem can potentially
pose two issues, one related to its tractability, and the other
related to its complexity. Very often, the worst case principle
leads to an intractable problem, since, as noted previously, the robust counterpart is an
optimization problem over an infinite set of constraints. Furthermore,
commonly employed approximation schemes usually increase the complexity of the
original problem by one degree, i.e., a linear program becomes an
SOCP and an SOCP transforms to an SDP.
In what follows, we propose a robust design which is merely based on
iteratively solving SOCPs, i.e., we attempt to minimize the
complexity of the robust version of the balancing problem. Interestingly enough, we note that the
SOCP-based scheme can also encompass a wide variety of uncertainty sets. In order to emphasize the
capability of the SOCP scheme to handle a variety of uncertainty sets, we will
not specify any particular norm to represent the uncertainty set.

We will arrive at a reduced complexity tractable robust scheme by
incorporating uncertainty and exploiting the structure of the SOC constraints in
\eqref{eq:ideal_SOCP}. To start with, we consider a relaxation of
\eqref{eq:worst_case_orig}, which is written as
\begin{subequations}\label{12e}
\begin{align}
&\underset{}{\operatorname{maximize}}& & t\notag\\
&\operatorname{subject\;to}\notag\\&&&
\hspace{-23.5mm}C_k\Re\Biggl(\bigg[{\vec{\hat
h}}_{b_k,k}+\sum_{i=1}^{l_{b_k,k}}\boldsymbol\delta
_{b_k,k}^i[{\vec{v}}_{b_k,k}]_i\bigg]{\vec m}_k\Biggr)-\left\|(z_{1,k}\; \ldots\;
z_{B,k}  \;\sigma)\trans \right\|_2\geq 0,\forall k,\notag\\&&&
\forall
{\vec{v}}_{b_k,k}:\|{\vec{v}}_{b_k,k}\|\leq
\rho_{b_k,k}\label{12ea}\\&&&\hspace{-23.5mm}z_{b,k}\!-\!\bigg\|{\vec
M}_b \herm \bigg[{\vec {\hat h}}_{b,k}+\sum_{i=
1}^{l_{b,k}}\boldsymbol\delta
_{b,k}^i[{\vec{v}}_{b,k}]_i\bigg]\herm
\bigg\|_2\geq 0,\forall b,k,\notag\\&&&\forall {\vec{v}}_{b,k}:\|{\vec{v}}_{b,k}\|\leq \rho_{b,k},\label{12eaa}\\
& & &\hspace{-12.5mm}z_{b,k}\geq 0, \forall b,k,\quad\|\mathrm{vec} ({\vec M}_b)\|_2\leq \sqrt{P_b}, \quad\forall b,\label{12ed}
\end{align}
\end{subequations}
where $C_k=\sqrt{1+\frac{\alpha_k}{t}}$, ${\vec
m}_k\in\mathbb{C}^{T\times 1},z_{b,k}\in \mathbb{R} ,{\vec
M}_b=[{\vec m}_{\mathcal{U}_b(1)}, \ldots, {\vec
m}_{\mathcal{U}_b(|\mathcal{U}_b|)}]$ are optimization variables and
we have removed the absolute value, and only consider the real part
of the left side of the constraints in \eqref{12ea}. Unlike the non-robust version of the
problem \eqref{eq:ideal_SOCP}, we cannot force the imaginary part of
$[{\vec{\hat{h}}}_{b_{k},k}+\sum_{i=1}^{l_{b_{k},k}}\boldsymbol{\delta}_{b_{k},k}^{i}[{\vec{v}}_{b_k,k}]_i]{\vec{m}}_{k}$
to zero for all channel error realizations. Since for a complex number $x$, $|x|\geq\Re(x)$,
a feasible point for \eqref{12ea} is also
feasible for the exact robust counterpart given in
\eqref{eq:worst_case_orig}. That is to say, to arrive at a reduced
complexity approach, we consider a conservative approximation of the
exact robust counterpart of the problem. For notational simplicity, when clear from context, we avoid mentioning the real operator $\Re(\cdot)$
explicitly from this point onwards.\par Now we make a key manipulation by substituting ${\vec{v}}_{b,k}=\bi{\uptheta}_{b,k}-\bi{\upphi}_{b,k}$
for all $b,k$, in \eqref{12eaa} and $\Re({\vec{v}}_{b_k,k})=\Re(\bi{\uptheta}_{b_k,k})-\Re(\bi{\upphi}_{b_k,k})$ such that
$\Re(\bi{\uptheta}_{b_k,k})\geq 0$ and $\Re(\bi{\upphi}_{b_k,k})\geq 0$ in \eqref{12ea}. After this we obtain a relaxation of \eqref{12e}
\begin{subequations}\label{12}
\begin{align}
&\underset{}{\operatorname{maximize}}& & t\notag\\
&\operatorname{subject\;to}\notag\\&&&
\hspace{-23.5mm}C_k\Biggl(\Re({\vec{\hat
h}}_{b_k,k}{\vec m}_k)+\sum_{i=1}^{l_{b_k,k}}\Re(\boldsymbol\delta_{b_k,k}^i{\vec m}_k)\biggl(\Re([\bi{\uptheta}_{b_k,k}]_i)-\Re([\bi{\upphi}_{b_k,k}]_i)\biggr)\Biggr)-\left\|(z_{1,k}\; \ldots\;
z_{B,k}  \;\sigma)\trans \right\|_2\geq 0,\forall k,\notag\\&&&
\forall
\bi{\uptheta}_{b_k,k},\bi{\upphi}_{b_k,k}:\|\Re(\bi{\uptheta}_{b_k,k})+\Re(\bi{\upphi}_{b_k,k})\|\leq
\rho_{b_k,k}^{\prime}\label{12a}\\&&&\hspace{0mm}z_{b,k}\!-\!\bigg\|{\vec
M}_b \herm \bigg[{\vec {\hat h}}_{b,k}+\sum_{i=
1}^{l_{b,k}}\boldsymbol\delta
_{b,k}^i([\bi{\uptheta}_{b,k}]_i-[\bi{\upphi}_{b,k}]_i)\bigg]\herm
\bigg\|_2\geq 0,\forall b,k,\notag\\&&&\forall \bi{\uptheta}_{b,k},\bi{\upphi}_{b,k}:\||\bi{\uptheta}_{b,k}|+|\bi{\upphi}_{b,k}|\|\leq \rho_{b,k}^{\prime},\label{12aa}\\
& & &\hspace{-12.5mm}z_{b,k}\geq 0, \forall b,k,\quad\|\mathrm{vec} ({\vec M}_b)\|_2\leq \sqrt{P_b}, \quad\forall b,\label{12d}
\end{align}
\end{subequations}
where
$[\bi{\uptheta}_{b_k,k}]_i,[\bi{\upphi}_{b_k,k}]_i$ denote the $i$th
components of $\bi{\uptheta}_{b_k,k},\bi{\upphi}_{b_k,k}$ for all
$b,k$, respectively, and for a vector ${\vec y}$ the symbol
${|{\vec y}|}$ represents that $[|{\vec y}|]_i=|[{\vec y}]_i|$ for all $i$. Another change
introduced in \eqref{12} is that for all $b,k$ we have replaced $\rho_{b_k,k}$ with $\rho^\prime_{b_k,k}$. The motivation for
this variation of the uncertainty set parameter $\rho_{b_k,k}$ becomes clear as we
outline the fact that splitting the uncertainty vector ${\vec v}_{b_k,k}$ into
a difference of two vectors and manipulating the left side of uncertainty sets as
done in \eqref{12} transforms the problem into \emph{nearly} a \emph{safe approximation} of its
original version while also rendering it \emph{tractability}.
\begin{remark}
As noted earlier, it appears difficult, if not impossible, to cast
the worst case robust counterpart \eqref{eq:worst_case_orig} into its exact equivalent tractable formulation. For example, in the
first approach based on SDP formulation, we had to drop the unit rank constraints to arrive at a tractable representation.
Naturally, to arrive at an SOCP representation of the problem, we have to resort to an approximation of the
original feasible set i.e.,
\begin{align}
\mathcal{O}_{orig}=\{\textrm{Optimization variables in \eqref{eq:worst_case_orig} such that all constraints in \eqref{eq:worst_case_orig} are satisfied}\}\label{OrigFea}
\end{align}
with its \emph{tractable subset} that may also include some additional analysis variables.
\end{remark}
In our case, it should be a feasible set for an SOC problem. By doing so we can ensure that
a solution for the approximation is definitely feasible for the original optimization program, and thus promises \emph{safety} in a sense that
we do not violate the original constraints. By just considering the real part of the left side of
constraints in \eqref{12ea}, we follow this strategy. It is worthwhile to note that the approximation used in \eqref{12a} may compromise safety.
However, our numerical investigations in Sec.~\ref{res} reveal that the safety of the proposed
SOCP procedure is almost guaranteed.  It is evident that by deriving a subset of the
original feasible set, we may have an overly conservative approximation. Hence, manipulating the parameter $\rho_{b,k}$ as $\rho^\prime_{b,k}$ may
provide some flexibility to overcome the conservativeness of the proposed approximation \cite{bertsimasCont}.

Next, let us first focus on the set of constraints in \eqref{12aa} and rewrite it in a form similar to the one
presented in \eqref{12eaa}
\begin{align}
z_{b,k}\!-\!\bigg\|{\vec
M}_b \herm \bigg[{\vec {\hat h}}_{b,k}+\sum_{i=
1}^{l_{b,k}}\boldsymbol\delta
_{b,k}^i[{\vec{v}}_{b,k}]_i\bigg]\herm
\bigg\|_2\geq 0,\forall b,k,\forall {\vec{v}}_{b,k}:\|{\vec{v}}_{b,k}\|\leq \rho_{b,k}^\prime\label{OrigCons1}.
\end{align}
It is worthy making an important observation now. A set of
optimization variables satisfies the constraints in \eqref{12aa} if and only if it satisfies the set of constraints in
\eqref{OrigCons1}.
For a given $b,k$, let us assume that $z_{b,k}$ and ${\vec
M}_b$ are infeasible in \eqref{12aa}, i.e., there exist $\bi{\uptheta}_{b,k},\bi{\upphi}_{b,k}$ and $\||\bi{\uptheta}_{b,k}|+|\bi{\upphi}_{b,k}|\|\leq \rho_{b,k}^\prime$
such that
\begin{align}
z_{b,k}-\bigg\|{\vec
M}_b \herm \bigg[{\vec {\hat h}}_{b,k}+\sum_{i=
1}^{l_{b,k}}\boldsymbol\delta
_{b,k}^i([\bi{\uptheta}_{b,k}]_i-[\bi{\upphi}_{b,k}]_i)\bigg]\herm
\bigg\|_2 < 0. \label{12aaVio}
\end{align}
Let $[{\vec{v}}_{b,k}]_i=[\bi{\uptheta}_{b,k}]_i-[\bi{\upphi}_{b,k}]_i$ for all $i$. Thus it is easy to see that $|[{\vec{v}}_{b,k}]_i|\leq |[\bi{\uptheta}_{b,k}]_i|+|[\bi{\upphi}_{b,k}]_i|$.
Therefore, we obtain $\Vert {\vec{v}}_{b,k}\Vert\leq \Vert |\bi{\uptheta}_{b,k}|+|\bi{\upphi}_{b,k}|\Vert\leq\rho_{b,k}^\prime$, and hence \eqref{OrigCons1} is also infeasible. Next, we assume conversely that for a given $b,k$, $z_{b,k}$ and ${\vec M}_b$ are infeasible in \eqref{OrigCons1}, i.e.,
\begin{align}
z_{b,k}\!-\!\bigg\|{\vec
M}_b \herm \bigg[{\vec {\hat h}}_{b,k}+\sum_{i=
1}^{l_{b,k}}\boldsymbol\delta
_{b,k}^i[{\vec{v}}_{b,k}]_i\bigg]\herm
\bigg\|_2< 0 \label{OrigConsVio}
\end{align}
for certain ${\vec{v}}_{b,k}$ such that $\|{\vec{v}}_{b,k}\|\leq\rho_{b,k}^\prime$. Let $[\bi{\uptheta}_{b,k}]_i=(1-\vartheta_{b,k})[{\vec{v}}_{b,k}]_i$ and $[\bi{\upphi}_{b,k}]_i=-\vartheta_{b,k}[{\vec{v}}_{b,k}]_i$, where $\vartheta_{b,k}\in[0,1]$.
With this substitution, it is seen that $[{\vec{v}}_{b,k}]_i=[\bi{\uptheta}_{b,k}]_i-[\bi{\upphi}_{b,k}]_i$. Similarly, these substitutions imply
$|[\bi{\uptheta}_{b,k}]_i|+|[\bi{\upphi}_{b,k}]_i|=|(1-\vartheta_{b,k})||[{\vec{v}}_{b,k}]_i|+|-\vartheta_{b,k}||[{\vec{v}}_{b,k}]_i|=|[{\vec{v}}_{b,k}]_i|$, and, thus, $\||\bi{\uptheta}_{b,k}|+|\bi{\upphi}_{b,k}|\|=\|{\vec{v}}_{b,k}\|\leq \rho_{b,k}^\prime$. Therefore, the variables $z_{b,k}$ and ${\vec M}_b$ are infeasible in \eqref{12aa} as well.
Hence, we conclude that the feasibility of the
constraints in \eqref{12aa} implies the feasibility of \eqref{OrigCons1}, and vice versa.

The main goal of the development so far is to
approximate \eqref{12a} and \eqref{12aa} by SOC constraints so that
the resulting robust counterparts in \eqref{12a}-\eqref{12d} can be
cast as an SOCP for fixed $t$. Since \eqref{12a} and \eqref{12aa}
have the same form, it is sufficient to concentrate on tackling the
more difficult set of constraints in \eqref{12aa}. We use the
concavity of the negative norm to bound \eqref{12aa} from below as
\begin{align}
&z_{b,k}\!-\!\big\|{\vec M}_b \herm {\vec {\hat
h}}_{b,k}\herm\big\|_2\!-\! \bigg\|{\vec M}_b \herm\!\bigg[\sum_{i=
1}^{l_{b,k}}\boldsymbol\delta
_{b,k}^i([\bi{\uptheta}_{b,k}]_i-[\bi{\upphi}_{b,k}]_i)\bigg]\herm\bigg\|_2\geq
0,\notag\\&\hspace{23 mm}\forall b,k, \forall
\bi{\uptheta}_{b,k},\bi{\upphi}_{b,k}:\||\bi{\uptheta}_{b,k}|+|\bi{\upphi}_{b,k}|\|\leq
\rho_{b,k}^{\prime}.\label{13a}
\end{align}
Again using the concavity argument, the left side of constraints in
\eqref{13a} can be further lower bounded as
\begin{align}
\label{14a}&z_{b,k}-\big\|{\vec M}_b \herm {\vec {\hat
h}}_{b,k}\herm\big\|_2+\sum_{i=1}^{l_{b,k}}\left[-\left\|
\{\boldsymbol\delta_{b,k}^i{\vec M}_b\}\herm
[\bi{\uptheta}_{b,k}]_i\right\|_2-\left\|\{\boldsymbol\delta_{b,k}^i{\vec
M}_b\}\herm(-[\bi{\upphi}_{b,k}]_i)\right\|_2\right]\geq 0,\forall
b,k\notag\\& \qquad\qquad\qquad\qquad\qquad\qquad
\forall \bi{\uptheta}_{b,k},\bi{\upphi}_{b,k}:\||\bi{\uptheta}_{b,k}|+|\bi{\upphi}_{b,k}|\|\leq \rho_{b,k}^{\prime}.
\end{align}
Reading the inequalities from \eqref{14a} backwards, and recalling the equivalence of \eqref{12aa} and \eqref{OrigCons1}, we
observe that a solution of \eqref{14a} is also feasible for \eqref{OrigCons1} or \eqref{12eaa}.
\begin{remark}
Before presenting a tractable formulation of the constraints in
\eqref{14a}, we again note that the optimal solution
 of the proposed SOCP relaxation ideally should also be feasible for the original worst case robust counterpart in \eqref{eq:worst_case_orig}. Therefore, if
$\mathcal{O}_{socp}$ represents the feasible set for the SOCP relaxation, then $\mathcal{O}_{socp}\subseteq\mathcal{O}_{orig}$ should hold. This will imply
both safety and tractability for the proposed SOCP approximation. Although the
transformations that will lead to an SOCP formulation for \eqref{14a} ensure both
these factors, the same cannot be observed for the constraint in \eqref{12a}. Nevertheless, as also noted above,
we will see in Sec. \ref{res} that the relation $\mathcal{O}_{socp}\subseteq\mathcal{O}_{orig}$ almost
remains valid at least for the cases considered. Being nearly a subset of the original problem, the proposed approximation can be rather
conservative, as also noted in \cite{bertsimasCont,bertsimas2}. To provide more flexibility in this regard, and as mentioned above, we make $\rho_{b,k}$ a design parameter and replace it
by $\rho_{b,k}^\prime$ in \eqref{12}. With the introduction of this maneuver, we may be able to improve the
achieved objective, albeit this may come at the cost of degradation in achieving it for given realizations of channel errors as we probe in the results section.\end{remark}\par
Let us define
\begin{align}
f_1({\vec M}_b,z_{b,k},{\vec {\hat h}}_{b,k})\triangleq
z_{b,k}-\big\|{\vec M}_b \herm {\vec {\hat
h}}_{b,k}\herm\big\|_2,\:\:f_2({\vec
M}_b,\boldsymbol\delta_{b,k}^i)\triangleq -
\left\|\{\boldsymbol\delta_{b,k}^i{\vec
M}_b\}\herm\right\|_2.\label{15}
\end{align}
We note that $f_2({\vec M}_b,\boldsymbol\delta_{b,k}^i)=f_2({\vec
M}_b,-\boldsymbol\delta_{b,k}^i)$. With the above definitions and using the fact that $\|k{\vec
m}\|_2=|k|\|{\vec m}\|_2$, the
constraints in \eqref{14a} can be equivalently written as
\begin{align}
&f_1({\vec M}_b,z_{b,k},{\vec {\hat
h}}_{b,k})+\displaystyle\min_{\||\bi{\uptheta}_{b,k}|+|\bi{\upphi}_{b,k}|\|\leq
\rho_{b,k}^{\prime}} \sum_{i=1}^{l_{b,k}}\bigg\{f_2({\vec
M}_b,\boldsymbol\delta_{b,k}^i)|[\bi{\uptheta}_{b,k}]_i|+f_2({\vec
M}_b,-\boldsymbol\delta_{b,k}^i)|[\bi{\upphi}_{b,k}]_i|\bigg\}\geq
0.\label{21}
\end{align}
The constraint in \eqref{21} can be cast into tractable form using
\cite[Theorem~1]{bertsimas2}, which is stated as:
\newtheorem{theorem}{Theorem}
\begin{theorem}\label{th1}
{\color{black} Working in the real domain, given a function $f({\vec x},{\vec U})$ that is concave
in data ${\vec U}$ for all given ${\vec x}$ and scales linearly with
the data}, we consider a constraint of the following form
\begin{align}
\displaystyle \min_{{\vec u}_1,{\vec u}_2\geq0:\|{\vec u}_1+{\vec u}_2\|\leq \omega}f({\vec
x},{\vec U}^n)+\sum_{j}\left[f({\vec x},\boldsymbol\delta^j)[{\vec u}_1]_j+f({\vec x},-\boldsymbol\delta^j)[{\vec
u}_2]_j\right]\geq 0\label{b1}
\end{align}
where ${\vec U}^n$ is the nominal part of the data,
$\boldsymbol\delta^j$ is a vector representing perturbation
direction in the $j$th component of the data and
{\color{black}${\vec u}_1$ and ${\vec u}_2$ are real vectors of
appropriate dimensions and the norm in \eqref{b1} satisfies the
property \cite[Eq.~6]{bertsimas2}
\begin{equation}\label{norm:prop}
   \|{\vec u}\|=\|{\vec |u|}\|
\end{equation}
where ${\vec |u|}=(|u_1|,\ldots,|u_{d}|)$}.
The constraint \eqref{b1} admits an equivalent representation of the form $f({\vec x},{\vec U}^n)\geq
\omega\|\boldsymbol \upgamma\|^\star$,
where $[\boldsymbol \upgamma]_j=\max\{-f({\vec x},\boldsymbol\delta^j),-f({\vec x},-\boldsymbol\delta^j)\}\geq 0$ and $\|\boldsymbol
\upgamma\|^\star\triangleq\displaystyle\max_{\|{\vec s}\|\leq
1}{\vec s}\trans\boldsymbol \upgamma$ is the dual norm of
$\boldsymbol \upgamma$.
\end{theorem}
\begin{IEEEproof}
The proof of the theorem is available in \cite[Theorem~1]{bertsimas2}. However,
for the sake of completeness and for demonstrating its applicability on \eqref{21} it is relegated to the Appendix.
\end{IEEEproof}

It should be emphasized that the norm in \eqref{b1} can be
arbitrary, as long as it satisfies \eqref{norm:prop}, meaning that the proposed
SOCP-based scheme presented next is applicable to a wide
variety and combinations of norms and thus uncertainty sets. For the special case of $l_2$
norm, the norms remain $l_2$ because
of the self dual property of the $l_2$ norm. Following similar steps
used to tackle the constraints in~\eqref{12aa},
we can easily see that the uncertain constraints in
\eqref{12a} can be cast in a form that is amenable to applying
Theorem~\ref{th1}. However, some important observations
should be re-stressed at this point. Although the constraints in \eqref{12a} are linear,
the solution of this approximation does not
necessarily imply \eqref{12ea}. This differs from the
previous scenario where the conversion of the constraints in \eqref{12eaa} to
\eqref{12aa} is safe. Therefore, obtaining a
safe, tractable and least possible conservative version of \eqref{12ea}
is left as an open question for future research.

With the aid of Theorem~\ref{th1}, the approximate robust
counterpart of the original problem can be written in the following
tractable form
\begin{subequations}\label{23App}
\begin{align}
&\underset{}{\operatorname{maximize}}& & t\notag\\
&\operatorname{subject\;to}& & C_k{\vec{\hat h}}_{b_k,k}{\vec
m}_k-\left\|\begin{pmatrix}z_{1,k} & \cdots & z_{B,k} &
\sigma\end{pmatrix}\trans\right\|_2\geq\rho_{b_k,k}^{\prime}
L_{b_k,k},\forall
k\label{23a}\\&&&C_k\boldsymbol\delta
_{b_k,k}^i{\vec m}_k+[{\vec q}_{b_k,k}]_q\geq 0,\forall
k,q=1,\ldots,l_{b_k,k}\label{23b}\\&&&\hspace{-5.8mm}-C_k\boldsymbol\delta
_{b_k,k}^i{\vec m}_k+[{\vec q}_{b_k,k}]_q\geq 0,\forall
k,q=1,\ldots,l_{b_k,k}, \:\: \|{\vec q}_{b_k,k}\|\leq
L_{b_k,k},\forall k \label{23c}\\&&& z_{b,k}-\big\|{\vec M}_b \herm {\vec {\hat
h}}_{b,k}\herm\big\|_2 \geq \rho_{b,k}^{\prime}\nu_{b,k},
\label{23d}\\&&& \hspace{-16.8mm}-
\left\|\{\boldsymbol\delta_{b,k}^i{\vec
M}_b\}\herm\right\|_2+[\bi{\upmu}_{b,k}]_i\geq 0,i=1\ldots
l_{b,k},\forall b,k,\:\: \|\bi{\upmu}_{b,k}\|\leq
\nu_{b,k},\forall b,k\label{23g}\\&&&\|\mathrm{vec} ({\vec M}_b)\|_2\leq \sqrt{P_b},
\quad\forall
b\label{23h}
\end{align}
\end{subequations}
where ${\vec m}_k\in\mathbb{C}^{T\times 1}, {\vec M}_b=[{\vec
m}_{\mathcal{U}_b(1)}, \ldots, {\vec
m}_{\mathcal{U}_b(|\mathcal{U}_b|)}], z_{b,k}\in\mathbb{R},
L_{b_k,k}\in\mathbb{R}, \nu_{b,k}\in\mathbb{R}, {\vec
q}_{b_k,k}\in\mathbb{C}^{l_{b_k,k}},
\bi{\upmu}_{b,k}\in\mathbb{C}^{l_{b,k}}$ are optimization variables.
The above optimization problem represents a tractable approximation,
in the form of SOCP in conjunction with bisection search, of the robust
counterpart of the problem under consideration. In the following, we provide some remarks regarding the tractability and reduced complexity of the proposed SOCP-based robust design.

\emph{Tractability}: We emphasize that while the SDP-based solution is only applicable to ellipsoidal uncertainty models, the SOCP-based approach is flexible enough to deal with
other types of uncertainty sets. For example, in certain situations, the errors in each of the individual terms of the
channel vector are bounded i.e., $|[{\vec v}_{b,k}]_i|\leq
\xi_{b,k}$ for all $b,k,i$.
This amounts to saying that $\|{\vec
v}_{b,k}\|_\infty\leq \xi_{b,k}$. In fact, in practical systems where each entry of ${\vec h}_{b,k}$ is quantized independently at the receiver and fed back to the corresponding transmitters, the interval uncertainty model is more appropriate \cite{Shenouda:Robust:2009}. Clearly, this uncertainty model
can be easily handled with the above approach since the
dual of the $l_\infty$ norm is well known \cite{nemirovski2}. In
other situations, it may happen that the entries of the uncertainty
vector are symmetrically random and bounded. In such scenarios it is
well known that the perturbation set can be represented as the
intersection of the $l_2$ and $l_\infty$ norms of ${\vec v}_{b,k}$
\cite{nemirovski3}. For this uncertainty model, it may be difficult, if not impossible,
to straightforwardly use worst case design philosophy. Hence, the SDP-based method
is not applicable and problem \eqref{eq:worst_case_orig} appears to
be intractable. However, the SOCP-based approach admits
tractability in the approximate solution of
the robust counterpart using the dual of the $l_2\cap l_\infty$ norm
\cite{nemirovski3}.

\emph{Complexity Reduction}:
The SOCP-based robust design also offers a great reduction in computational complexity compared to the SDP-based method. In what follows, we give a complexity comparison of the SDP- and SOCP-based solutions for the special case where ${\vec A}_{b,k}={\vec I}_T$ for all
$b,k$, which is commonly considered in the related works.\footnote{The same arguments in this part also apply to case where the
entries of channel vectors undergo independent perturbations.} First, let us focus on the equivalent representation
obtained using Theorem~\ref{th1} and explore it by considering any
robust equivalent constraint (without loss of generality) from
\eqref{23g}. We note that under this setting each entry
of a channel can be written as $[{\vec h}_{b,k}]_m=[{\vec {\hat
h}}_{b,k}]_m+[{\boldsymbol \delta}_{b,k}]_m [{\vec v}_{b,k}]_m,\; 1\leq
m\leq T$, where ${\vec v}_{b,k}$ belongs to the uncertainty set defined
in \eqref{s1}.
The vector ${\boldsymbol
\upgamma}_{b,k}$ corresponding to the equivalent formulation of
Theorem~\ref{th1} becomes
\begin{align}
{\boldsymbol \upgamma}_{b,k}\trans=[|[{\vec
m}_{\mathcal{U}_{b}(1)}]_m[{\boldsymbol
\delta}_{b,k}]_m|,\dots,|[{\vec
m}_{\mathcal{U}_{b}(|\mathcal{U}_{b}|)}]_m[{\boldsymbol
\delta}_{b,k}]_m|],\forall b,k. \label{red2}
\end{align}
With this type of ${\boldsymbol \upgamma}_{b,k}$ it has been shown
in \cite{bertsimas2} that for ellipsoidal uncertainty set, instead of
having multiple additional constraints of the type mentioned above,
we can stack all corresponding variables into one SOC constraint,
$\|\bi{\upmu}_{b,k}\|_2\leq \nu_{b,k}$, and one variable
$\nu_{b,k}$ for all $b,k$. Similarly, the constraints
involving user $(b_k,k)$, shown  in \eqref{23b}-\eqref{23c}, can be greatly simplified.

To provide a complexity comparison, we base our discussion on the simplification noted above and focus on an arbitrary bisection step.
According to \eqref{23a}-\eqref{23h}, the number of real optimization variables per bisection iteration of the SOCP-based robust design
is $4TBK+2BK+2KT+K$. More specifically, there are $BK$ constraints
of real dimension $(2TK_b+1)$ that occur thrice including the power
constraint. Again using the above mentioned simplification, we obtain two constraints of real dimensions $B+2$ and $T+1$ that are
$K$ in number. Combining all these the worst case per iteration
complexity  of the SOCP approach approximates as
%$\mathcal{O}((4TBK+2BK+2KT+K)^2\{K(B+2)+K(T+1)+3BK(2TK_b+1)\})$ \cite{vanden}.
$\mathcal{O}\bigl(K_b(TBK)^{3}\bigr)$ \cite{vanden,lobo}.
The per iteration complexity of
the SDP-based method is found to be $\mathcal{O}((KBT)^6)$
\cite{luoC,vanden}, which is clearly higher than the SOCP counterpart. Further to this,
based on \cite{lobo,vanden}, the worst case estimate of the number of iterations needed to arrive at a numerically acceptable value
of the SOCP-based design is \textcolor{black}{$\mathcal{O}(\sqrt{KB})$}. As similar calculation reveals that such an estimate for the SDP-based method
results in a higher value of \textcolor{black}{$\mathcal{O}(\sqrt{KTB})$} on account of its dependence on the size of the matrix inequalities. A more
detailed exploration that compares run times of the proposed approaches with different solvers is given in Sec. \ref{res}.
\vspace{0mm}
\section{Numerical Results}\label{res}
\begin{figure}
\centering \subfigure[Average worst-case SINR versus
$\rho$.]{\label{fig:worstSINR}\includegraphics{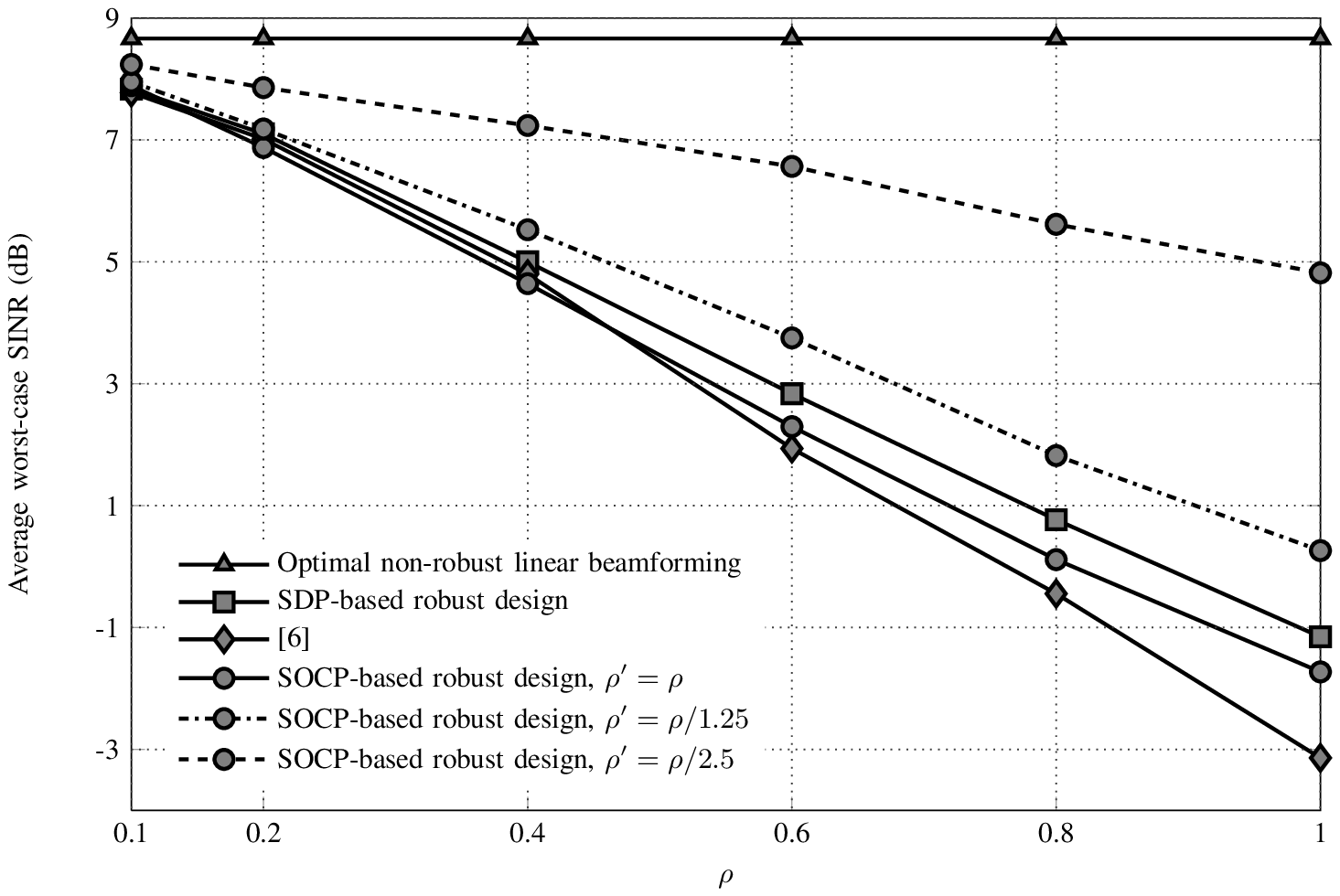}
}\\ \vspace{10pt}\subfigure[Variation of $\mathrm{PE}$ with $\rho$ for the
proposed SOCP-based robust design.]{\label{tab:PE:2norm}\footnotesize
\begin{tabular}{c||c|c|c|c|c|c} \hline  $\rho$  & $0.1$  & $0.2$  & $0.4$
& $0.6$  & $0.8$  & $1$\tabularnewline \hline  \hline $\mathrm{PE}$ non-robust  & $0$  & $0$  & $0$  & $0$  & $0$  &
$0$\tabularnewline \hline
$\mathrm{PE}$ SDP  & $1$  & $1$  & $1$  & $1$  & $1$  &
$1$\tabularnewline \hline $\mathrm{PE}$
\cite{tajer}  & $1$  & $1$  & $1$  & $1$  & $1$  &
$1$\tabularnewline \hline  $\mathrm{PE}$
($\rho^{\prime}=\rho$)  & $1$  & $1$  & $1$  & $1$  & $1$  &
$1$\tabularnewline \hline  $\mathrm{PE}$ ($\rho^{\prime}=\rho/1.25$)  &
$0.99$ & $1$ & $0.99$ & $0.99$ & $0.98$ & $0.99$\tabularnewline \hline
$\mathrm{PE}$ ($\rho^{\prime}=\rho/2.5$)  & $0.81$  & $0.87$  & $0.82$  &
$0.75$  & $0.80$  & $0.73$\tabularnewline \hline  \multicolumn{1}{c}{} &
\multicolumn{1}{c}{} & \multicolumn{1}{c}{} & \multicolumn{1}{c}{} &
\multicolumn{1}{c}{} & \multicolumn{1}{c}{} &
\multicolumn{1}{c}{}\tabularnewline \end{tabular} }
\caption{Average worst-case SINR versus $\rho$ for different approaches
where
channel uncertainties are bounded by $l_{2}$-norm. The value of power
for both BSs is taken as $5$ dB. The number of transmit antennas
at each BS is $T=8$. The total number of users is $K=4$ ($2$ users
per base station). }
\label{fig:SINRwithrho2norm}
\end{figure}

\begin{figure}
\centering{\includegraphics{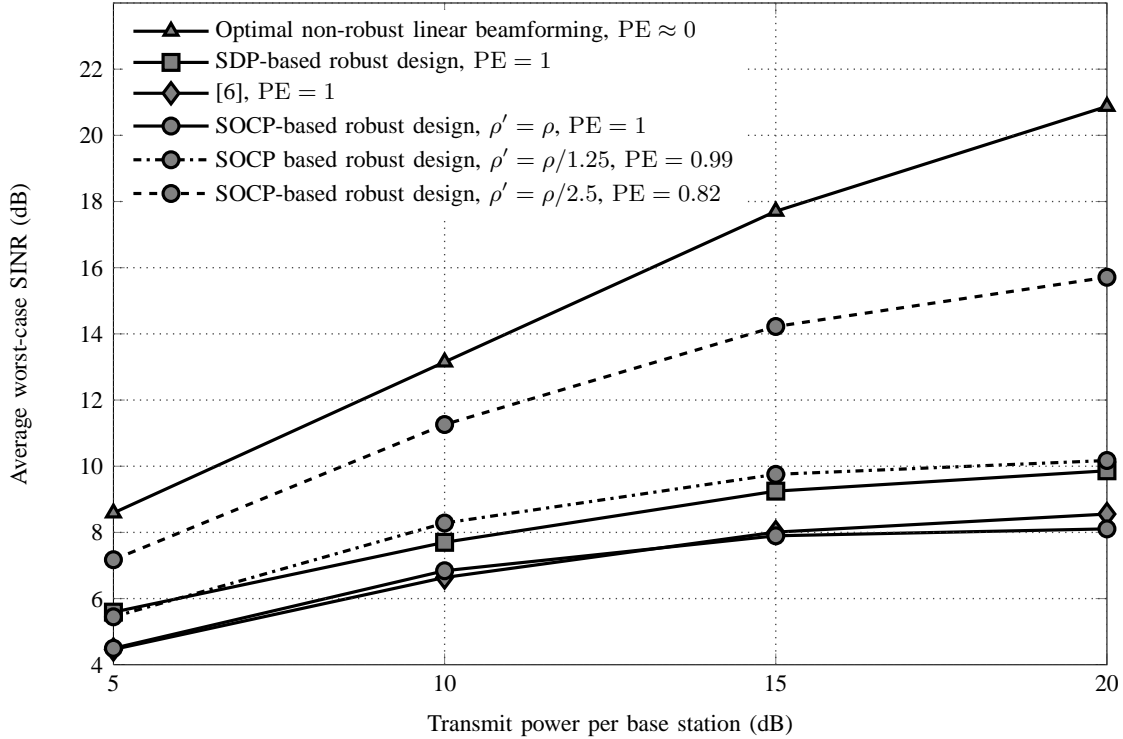}}
\caption{Comparison of the worst-case SINR of non-robust, SOCP-, SDP-based
design, and the approach of \cite{tajer} as a function of the transmit
power of the BSs for $\rho=0.4$. The performances of the SOCP-based
method are shown for three cases where $\rho^{\prime}=\rho$,
$\rho^{\prime}=\rho/1.25$
and $\rho^{\prime}=\rho/2.5$. }
\label{fig:worstSINR2normwithpower}
\end{figure}

\begin{figure}
\centering\subfigure[Variation of average worst-case SINR versus $\rho$ for
$l_{\infty}$-norm.]{\includegraphics{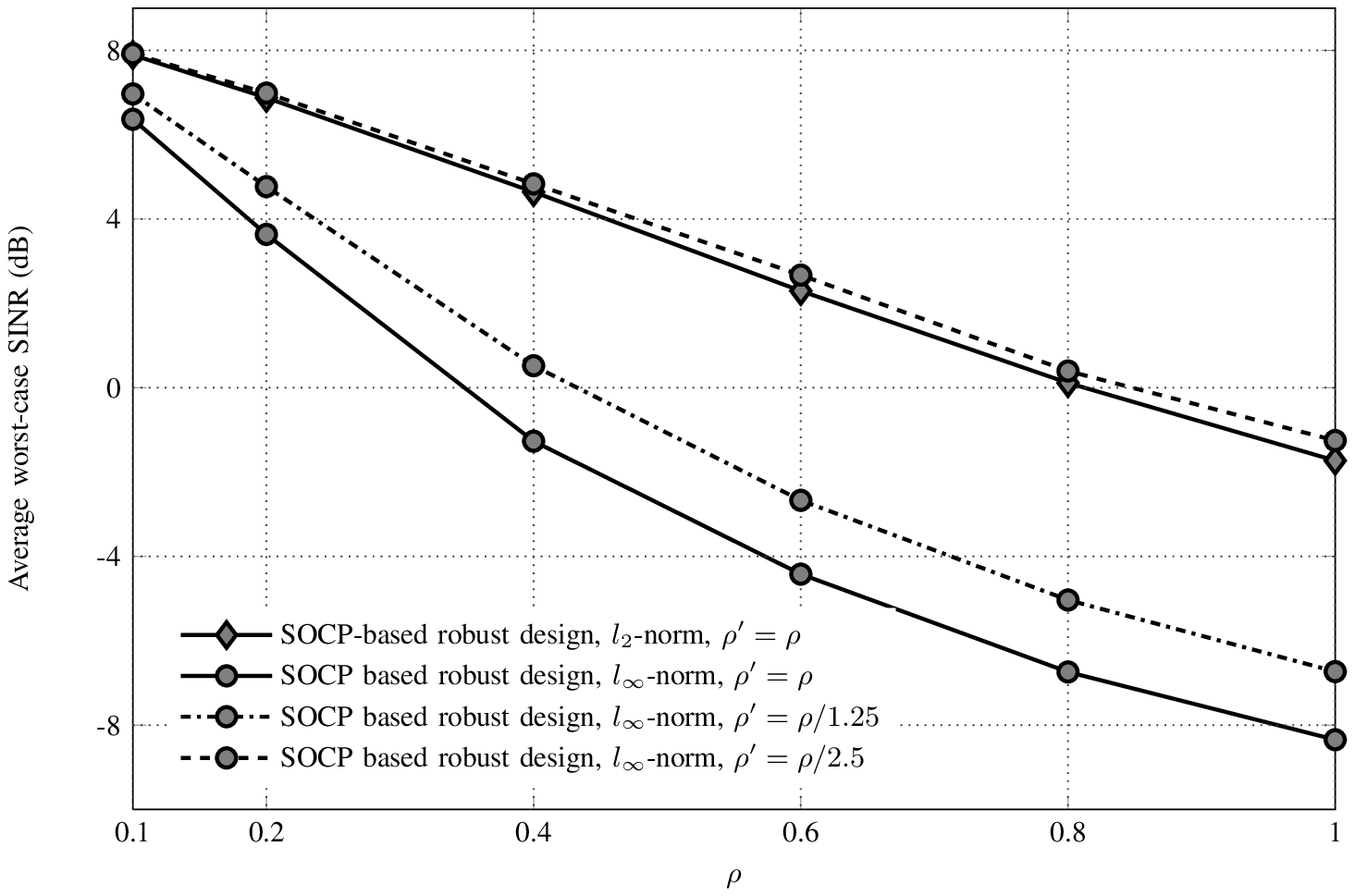}}\\
\vspace{10pt} \subfigure[Variation of $\mathrm{PE}$ with $\rho$ for
$l_{\infty}$-norm for the proposed SOCP-based robust
approach.]{\label{tab:PE:inftynorm}\footnotesize
\begin{tabular}{c||c|c|c|c|c|c} \hline  $\rho$  & $0.1$  & $0.2$  & $0.4$
& $0.6$  & $0.8$  & $1$\tabularnewline \hline  \hline  $\mathrm{PE}$
($\rho^{\prime}=\rho$)  & $1$  & $1$  & $1$  & $1$  & $0.96$  &
$0.98$\tabularnewline \hline  $\mathrm{PE}$ ($\rho^{\prime}=\rho/1.25$)  &
$1$ & $1$ & $0.99$ & $0.99$ & $0.95$ & $0.97$\tabularnewline \hline
$\mathrm{PE}$ ($\rho^{\prime}=\rho/2$)  & $1$ & $0.97$ & $0.96$ & $0.95$ &
$0.95$ & $0.91$\tabularnewline \hline  $\mathrm{PE}$
($\rho^{\prime}=\rho/2.5$)  & $0.8$  & $0.88$  & $0.81$  & $0.81$  & $0.81$
& $0.80$\tabularnewline \hline  \multicolumn{1}{c}{} & \multicolumn{1}{c}{}
& \multicolumn{1}{c}{} & \multicolumn{1}{c}{} & \multicolumn{1}{c}{} &
\multicolumn{1}{c}{} & \multicolumn{1}{c}{}\tabularnewline \end{tabular}}
\caption{Variation of average worst-case SINR versus $\rho$ for
$l_{\infty}$-norm
(i.e., {}``box'' uncertainty). The value of power for both BSs has
been taken as $5$ dB.The number of transmit antennas at each BS is
$T=8$. The total number of users is $K=4$.}
\label{fig:worstSINRinftynormwithrho}
\end{figure}

\begin{figure}
\centering{\includegraphics{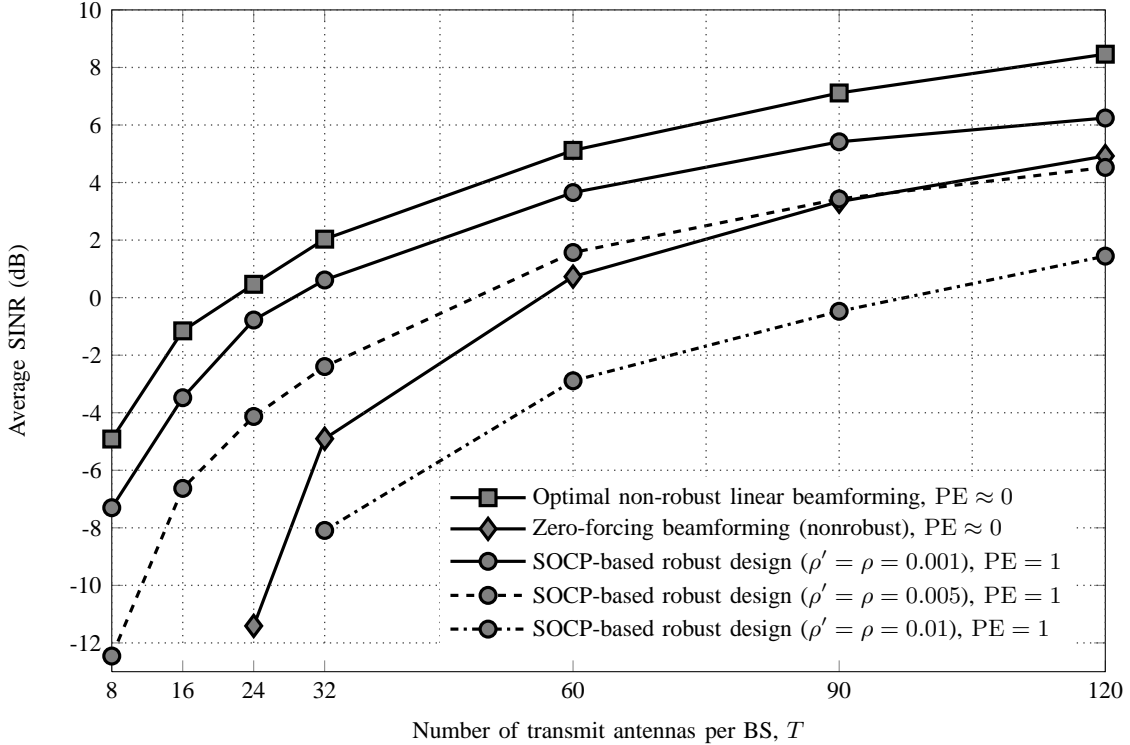}}
\caption{Variation of average balanced SINR versus the number of transmit
antennas
per BS, $T$. For robust designs, channel uncertainties are are bounded
in a ball of radius $\rho$. The value of the normalized power for both BSs is
taken as $20$ dB. The total number of users is $K=20$ ($10$ users
per cell).}
\label{fig:SINR:massiveMIMO}
\end{figure}

In order to compare the performance of the proposed approaches we
report results of numerical simulations in this section. For all simulation
setups, we consider a system of two cells ($B=2)$, while the number
of users per cell is mentioned for individual numerical experiments. The channel vector from
the BS $b$ to user $k$ is given by
$\mathbf{h}_{b,k}=\sqrt{\kappa_{b,k}}\tilde{\mathbf{h}}_{b,k}$
where $\kappa_{b,k}$ represents both the path loss and the shadow fading and
$\tilde{\mathbf{h}}_{b,k}$ follows $\mathcal{CN}(0,{\vec I})$. In Figs.
\ref{fig:SINRwithrho2norm}-\ref{fig:worstSINRinftynormwithrho} to
follow, we only consider the small-scale fading (i.e., $\kappa_{b,k}=1$
for all $b$ and $k$). These setups can be considered to correspond
to the worst-case scenario where all users are at the cell edge. A
more realistic channel model where large-scale fading is taken into
account is investigated in Fig. \ref{fig:SINR:massiveMIMO} for a
massive MISO system. All noise variances are taken as unity and the
transmit power is normalized with respect to the noise variance. For the sake of
simplicity, but without compromising generality, we take $\alpha_{k}=1$
for all $k$, ${\vec{A}}_{b,k}={\vec{I}}_{T}$, and $\rho_{b,k}=\rho$
for all $b,k$. Unless otherwise mentioned, the error vectors are assumed
to lie in a hypersphere of radius $\rho$. We evaluate the performance
of the three approaches in terms of the worst-case SINR (i.e.,
the objective obtained at the end of the bisection procedure when
solving \eqref{11nm} and \eqref{23App}) and the probability of exceeding the worst-case
SINR which is referred to as $\mathrm{PE}$ from now on. For the simulation setup considered in this paper, the SDP-based
approach is numerically found to produce precoding matrices close
to rank-1 matrices.

Fig. \ref{fig:SINRwithrho2norm} plots the average worst-case SINRs
(over 200 realizations of the nominal channels $\hat{\mathbf{h}}_{b,k}$)
versus the radius of the uncertainty sets, $\rho$, for all approaches.
In this simulation setup, we only consider the small-scale fading
(i.e., $\kappa_{b,k}=1$ for all $b$ and $k$). The number of users
in each BS is $2$, i.e., there are $K=4$ users in total. It is
seen in Fig. \ref{fig:worstSINR} that the SOCP-based solution gives the worst-case
SINR close to that of the SDP-based approach and slightly higher than
that of \cite{tajer} when $\rho^{\prime}=\rho$. As
mentioned earlier, by taking $\rho^{\prime}$ as a design parameter,
we can achieve a trade-off between the worst-case SINR and the resulting
$\mathrm{PE}$. It can be seen from Fig. \ref{fig:SINRwithrho2norm}
that the worst-case SINR of the SOCP-based solution is improved
when we take $\rho^{\prime}=\rho/2.5$, but this implies reduced
$\mathrm{PE}$
as indicated in table \ref{tab:PE:2norm}. The values of $\mathrm{PE}$
given in table \ref{tab:PE:2norm} are obtained with $10^{6}$ sets
of channel errors that are uniformly distributed in the ball of $\rho$
using the toolbox of \cite{RACT}. As expected, the non-robust approach delivers the
maximum SINR and virtually zero $\mathrm{PE}$ in all cases.

\begin{table}[h]
\caption{The average run time (in seconds) versus the number of transmit
antennas, $T$,
at each BS for the robust designs. The number of BSs is $B=2$,
each serving $10$ users. The bisection procedure terminates when the
difference between the objective values of two bisection steps,
$\epsilon\leq10^{-2}$.}
\label{runtime}\centering \footnotesize
\begin{tabular}{c||c|c|c|c|c|c|c|c|c}
\hline
Antennas  & $8$  & $12$  & $16$  & $50$  & $100$  & $200$  & $300$  & $400$  &
$500$\tabularnewline
\hline
\hline
SDP-based design (SDPT3) (sec)  & $96.48$ & $477.55$ & $5620.3397$ & $\times$  &
$\times$  & $\times$  & $\times$  & $\times$  & $\times$\tabularnewline
\hline
SDP-based design (SeDuMi) (sec)  & $31.44$ & $162.68$ & $684.57$ & $\times$  &
$\times$  & $\times$  & $\times$  & $\times$  & $\times$\tabularnewline
\hline
\cite{tajer} (SDPT3) (sec) & $88.34$ & $130.83$ & $240.33$ & $\times$  & $\times$
& $\times$ & $\times$ & $\times$ & $\times$\tabularnewline
\hline
\cite{tajer} (SeDuMi) (sec) & $48.01$ & $61.09$ & $156.78$ & $\times$  & $\times$
& $\times$ & $\times$ & $\times$ & $\times$\tabularnewline
\hline
SOCP-based design (SDPT3) (sec) & $1.63$ & $4.04$ & $4.07$ & $28.09$ & $99.23$ &
$285.74$ & $-$ & $-$ & $-$\tabularnewline
\hline
SOCP-based design (SeDuMi) (sec) & $0.66$ & $1.09$ & $1.23$ & $21.92$ & $51.31$ &
$149.68$ & $-$ & $-$ & $-$\tabularnewline
\hline
SOCP-based design (GUROBI) (sec) & $1.08$ & $1.95$ & $2.14$ & $12.02$ & $23.66$ &
$44.91$ & $47.37$ & $67.46$ & 90.72\tabularnewline
\hline
\end{tabular}
\end{table}

Although the SDP formulations can offer better worst-case SINR, they are
not practically useful for large-scale antenna systems especially from the complexity
perspective. Another disadvantage of the SDP approach is its inability to handle various uncertainty sets and limited
choice of solvers compared to those for SOCP based solutions. The
flexibility in choosing a solver is important because a general purpose
convex programing solver may not be efficient for all problems. We
compare the simulation time of the SDP and SOCP-based robust designs
using YALMIP \cite{YALMIP} with two widely used conic programming
solvers (SeDuMi \cite{sedumi} and SDPT3 \cite{sdpt3}). Note that the proposed SOCP-based method allows us to make
use of GUROBI \cite{GUROBI} as a solver as well which is claimed to be very
efficient for detecting feasibility of large-scale SOCPs. For the
robust SOCP-based design, we use the simplified representation in
\eqref{red2}. In Table \ref{runtime}, we show the average run time (in seconds)
of all robust approaches as a function of the number
of transmit antennas, $T$, for solving the corresponding optimization
problem. The bisection procedure terminates if feasibility is detected
and the relative difference $\epsilon$ of the objectives between two bisection
steps is less than or equal to $10^{-2}$. The lower threshold of
the bisection algorithm is set to $0$, while the upper one is equal
to the balanced SINR obtained from the non-robust design. The codes
are executed on a 64-bit desktop that supports 8 Gbyte RAM and
Intel CORE i7.
For both solvers, it can be clearly seen that the SOCP-based design
requires a lower run time and the difference is considerable as the
number of transmit antennas $T$ increases. This observation matches
with the theory presented in the subsection on reduced complexity
in Sec. \ref{Red_Comp}. Moreover, we notice that the SDP-based methods
are not capable of producing a solution when $T\geq50$ due to
lack of memory (denoted by a cross mark {}``$\times$''
in Table \ref{runtime}). When $T\geq300$, SeDuMi and SDPT3 are not
suitable solvers for the SOCP-based method since they are not able
to produce a solution even after several hours (denoted by {}``$-$''
in Table \ref{runtime}). We have observed that GUROBI is the most
efficient solver for the SOCP-based method in particular for large-scale antenna
array systems.

In Fig. \ref{fig:worstSINR2normwithpower}, the average worst-case
balanced SINR (again over 200 realizations of the nominal channels
$\hat{\mathbf{h}}_{b,k}$)
is plotted with the transmit power per BS, $P$, for different approaches.
We note that the SOCP-based approach performs nearly as good as the
SDP one. The reduced minimum SINR of \cite{tajer} can
be probably attributed to the fact that it is completely a conservative
approximation of the robust counterpart. Recall that, in our proposed
SOCP-based design, we can control the degree of conservatism of the
design by finding proper value of $\rho^{\prime}$. The values of
$\mathrm{PE}$ for three approaches are also provided in Fig.
\ref{fig:worstSINR2normwithpower}.
Further, being oblivious to channel error vectors, the non robust
design delivers the best worst-case SINR. Nonetheless, as expected and seen previously,
this comes at the cost of unacceptably low $\mathrm{PE}$, i.e.,
$\mathrm{PE}\approx0$.
It is found that the SOCP-based method gives $\mathrm{PE}=0.99$ for
$\rho^{\prime}=\rho/1.25$, while the approach of \cite{tajer} and
the SDP-based solution both produce $\mathrm{PE}=1.0$. The value
of $\mathrm{PE}$ for the SOCP-based design is reduced to $0.82$
as we set $\rho^{\prime}=\rho/2.5$. Interestingly, this decrease
in PE is accompanied by a corresponding increase in the worst-case
SINR, thereby providing a tradeoff between the two parameters. We
note that the trend of values of $\mathrm{PE}$ is observed to be
typical for the range of transmit power considered in Fig.
\ref{fig:worstSINR2normwithpower}.

In Fig. \ref{fig:worstSINRinftynormwithrho} we evaluate performance
of robust beamforming for SINR balancing where the errors in elements
of channel vectors are bounded within a (multi-dimensional) box of
size $\rho$, i.e., $|[{\vec{v}}_{b,k}]_{i}|\leq\rho$ for all
$i$. This is equivalent to saying that
$\|{\vec{v}}_{b,k}\|_{\infty}\leq\rho$.
For this case, we note that the SDP formulations and the approximations
used in \cite{tajer} are not applicable. The curves in Fig.
\ref{fig:worstSINRinftynormwithrho}
have been obtained by noting the fact that the dual of $l_{\infty}$-norm
is $l_{1}$-norm. It is seen that, with box uncertainty (Fig.
\ref{fig:worstSINRinftynormwithrho}),
the worst-case SINR is lower than that in the case of ellipsoidal
uncertainty for the same $\rho$. This can be explained as follows.
We note that for a vector ${\vec{v}}$,
$\|{\vec{v}}\|_{1}\geq\|{\vec{v}}\|_{2}\geq\|{\vec{v}}\|_{\infty}$,
which means that for the same $\rho$, the $l_{\infty}$-norm defines
a smaller feasible set in \eqref{23App} compared to the $l_{2}$-norm. Thus the worst-case
SINR for the $l_{\infty}$-norm uncertainty is lower than that of the
$l_{2}$-norm. However, when errors are uniformly distributed in a box, we note
in the table given in Fig. \ref{tab:PE:inftynorm} a slight degradation in PE when $\rho^\prime=\rho$.
This stems from the fact that the proposed approach is not
guaranteed to be safe as discussed earlier in the paper.

Finally, in Fig. \ref{fig:SINR:massiveMIMO} we investigate how the balanced SINR of all users scales
with the number of transmit antennas. In particular, we consider a
system of two cells, each serving $10$ users. The users are uniformly
distributed in the cell and are not allowed to be closer to the BS by more than
$d_{0}=100$ meters \cite{Hien:MassiveMIMO:2013}. We also assume
that the cell diameter (to a vertex) is $1000$ meters. The large-scale
fading coefficient is modeled as
$\kappa_{b,k}=\beta_{b,k}(d_{b,k}/d_{0})^{-\nu}$
where $\beta_{b,k}$ accounts for shadow fading assumed to follow
a log-normal distribution with standard deviation
$\sigma_{\textrm{shadow}}$,
$\nu$ is the path loss exponent, and $d_{b,k}$ is the distance between
BS $b$ and user $k$. In Fig. \ref{fig:SINR:massiveMIMO}, we choose
$\sigma_{\textrm{shadow}}=8$ dB and $\nu=3.8$ as in
\cite{Hien:MassiveMIMO:2013}.
In Fig. \ref{fig:SINR:massiveMIMO}, the performance of zero-forcing
beamforming (ZF-BF) scheme is also included for comparison. In ZF-BF,
multiuser interference for each user is forced to zero, i.e.,
$\mathbf{h}_{b_{i},j}\mathbf{m}_{i}=0$
for all $j\neq i$ \cite{Nam:UserSelectionBD:2012,Nam:beamdesign:ZFDPC:2012}.
In this way, problem \eqref{eq:ideal_SOCP} is simplified as
\begin{eqnarray}
\underset{(t,\mathbf{m}_{k})\in\mathcal{F}}{\operatorname{maximize}} &  & t\label{eq:zf}
\end{eqnarray}
where
$\mathcal{F}\triangleq\{(t,\mathbf{m}_{k})\bigl|\mathbf{h}_{b_{i},j}\mathbf{
m}_{i}=0,\,\forall i\neq
j,\;\|\mathbf{M}_{b}\|_{2}\leq\sqrt{P_{b}},\,\forall
b,\;\sigma\frac{t}{\alpha_{k}}\leq\Re(\mathbf{h}_{b_{k},k}\mathbf{m}_{k}),\,
\forall k\}$
which is convex. Consequently, problem
\eqref{eq:zf}
is jointly convex in $t$ and $\mathbf{m}_{k}$. Using the null-space
technique as devised in \cite{Nam:UserSelectionBD:2012,Nam:beamdesign:ZFDPC:2012},
we can further remove the ZF constraints
$\mathbf{h}_{b_{i},j}\mathbf{m}_{i}=0,\;\forall i\neq j$
in $\mathcal{F}$ without loss of optimality as follows. Let
$\bar{\mathbf{M}}_{k}=[\mathbf{h}_{b_{1,k}}\trans\;\mathbf{h}_{b_{2,k}}\trans\,\cdots\,\mathbf{h}_{b_{k-1,k}}\trans\;\mathbf{h}_{b_{k+1,k}}\trans\,\cdots\,\mathbf{h}_{b_{K,k}}\trans]\trans\in\mathbb{C}^{(K-1)\times T}$
and $\mathbf{G}_{k}\in\mathbb{C}^{T\times(T-K+1)}$ be a matrix of
orthogonal columns that span the null space of $\bar{\mathbf{M}}_{k}$.%
\footnote{For the ZF-BF scheme to be feasible, i.e.,
$\dim(\mathbf{G}_{k})>0$,
we must have $T\geq K$.%
} Then, to satisfy the ZF constraints, we can write
$\mathbf{m}_{k}=\mathbf{G}_{k}\tilde{\mathbf{m}}_{k}$.
The problem \eqref{eq:zf} is thus further equivalent to
\begin{eqnarray}
\underset{(t,\tilde{\mathbf{m}}_{k})\in\tilde{\mathcal{F}}}{\operatorname{maximize}} &  &
t\label{eq:zf:reform}
\end{eqnarray}
where
$\tilde{\mathbf{h}}_{b_{k},k}\triangleq\mathbf{h}_{b_{k},k}\mathbf{G}_{k}$
and
$\tilde{\mathbf{M}}_{b}\triangleq[\tilde{\mathbf{m}}_{U_{b}(1)},...,\tilde{\mathbf{m}}_{U_{b}(|U_{b}|)}]$
and
$\tilde{\mathcal{F}}\triangleq\{\tilde{\mathbf{m}}_{k}\bigl|\|(\tilde{\mathbf{M}}_{b})\|_{2}\leq\sqrt{P_{b}}\,\forall b,\;\sigma\frac{t}{\alpha_{k}}\leq\Re(\tilde{\mathbf{h}}_{b_{k},k}\tilde{\mathbf{m}}_{k}),\,\forall k\}$.
The advantage of the ZF-BF scheme is that we can avoid the bisection
procedure that must be carried out to solve \eqref{eq:ideal_SOCP} for optimal linear
beamforming. However, even for perfect CSI, the performance of ZF-BF is still far away from
that of optimal linear beamforming as shown in Fig.
\ref{fig:SINR:massiveMIMO}.
For example, when $T=120$, a gap of about $4$ dB is observed between ZF-BF
and general linear beamforming. These non-robust designs are sensitive
to channel errors as shown by the fact that $\mathrm{PE}\approx0$ for both these cases.
The benefit of using large-scale antenna systems is seen in
Fig. \ref{fig:SINR:massiveMIMO}, when a gain of $13$ dB is observed as the
number of transmit antennas is increased from $T=8$ to $T=120$ for
optimal linear beamforming with perfect CSI. A similar conclusion also applies to the
robust designs when $\rho$ is taken as $0.001$, $0.005$ and $0.01$.
We note that these values of $\rho$ are comparable to the average
channel gains of $\mathbf{h}_{b,k}$ taking into account the effect
of shadowing and path loss. Therefore, owing to the conservative nature of robust designs
it is not possible to achieve nontrivial SINRs for higher values of $\rho$.
\section{Conclusion}\label{con}
We have studied the design of beamformers that balance the SINR of
users in a multicell downlink system in the
presence of channel uncertainties.
Norm bounded channel uncertainty model is used. As a first approach
to solving the problem in this scenario, we present an
$\mathcal{S}$-lemma based approximate solution in which the
beamformers are obtained by solving an SDP in conjunction with
bisection search. Later, by exploiting various properties of the
functions involved in the problem, we present a solution in which
robust beamformers are solutions to an SOCP-based formulation. We
show that in addition to being capable of handling different
uncertainty sets, the SOCP-based
approximation exhibits a much reduced complexity solution. We have tested the performance of
proposed approaches for the recently conceived massive antenna systems, and have determined that the SOCP
approach outperforms the SDP-based solution from computational cost perspective. Finally,
we have also shown that the
reduced complexity SOCP-based approach yields a balanced SINR and the probability of achieving it which is
comparable with the SDP approach.
\section*{Acknowledgment}
The authors would like to thank Prof. Johan Löfberg of the Linköping University,
Sweden for his helpful advice regarding the numerical results
presented in the paper.
\appendix[Proof of Theorem \ref{th1}]
We will follow the notation in \eqref{21} and obtain a tractable version of this constraint
by adapting the arguments developed in \cite{bertsimas2}. Let $O_1$
and $O_2$ be the optimal solutions of

\begin{subequations}\label{A1}
\begin{align}
&\underset{}{\operatorname{max}}& & {\vec a}_1\trans{\vec v}_1+{\vec a}_2\trans{\vec v}_2\label{A1a}\\
&\operatorname{subject\;to}&&
\|{\vec v}_1+{\vec v}_2\|\leq \vartheta\label{A1b}\\&&&
{\vec v}_1\geq 0,{\vec v}_2\geq 0\label{A1c}.
\end{align}
\end{subequations}
and
\begin{subequations}\label{A2}
\begin{align}
&\underset{}{\operatorname{max}}& & \sum_{i\in\mathcal{I}}\max\{[{\vec a}_1]_i,[{\vec a}_2]_i,0\}[{\vec v}_3]_i\label{A2a}\\
&\operatorname{subject\;to}&&
\|{\vec v}_3\|\leq \vartheta.\label{A2b}
\end{align}
\end{subequations}
respectively. It is shown in \cite{bertsimas2} that $O_1=O_2$. Now consider the following set of relations that can be obtained from \eqref{21}, i.e.,
\begin{subequations}\label{A3}
\begin{align}
f_1({\vec M}_b,z_{b,k},{\vec {\hat
h}}_{b,k})&\geq-\displaystyle\min_{\||\bi{\uptheta}_{b,k}|+|\bi{\upphi}_{b,k}|\|\leq
\rho_{b,k}^{\prime}} \sum_{i=1}^{l_{b,k}}\bigg\{f_2({\vec
M}_b,\boldsymbol\delta_{b,k}^i)|[\bi{\uptheta}_{b,k}]_i|+f_2({\vec
M}_b,-\boldsymbol\delta_{b,k}^i)|[\bi{\upphi}_{b,k}]_i|\bigg\}\label{A3a}\\&
=\displaystyle\max_{\||\bi{\uptheta}_{b,k}|+|\bi{\upphi}_{b,k}|\|\leq
\rho_{b,k}^{\prime}} \sum_{i=1}^{l_{b,k}}\bigg\{-f_2({\vec
M}_b,\boldsymbol\delta_{b,k}^i)|[\bi{\uptheta}_{b,k}]_i|-f_2({\vec
M}_b,-\boldsymbol\delta_{b,k}^i)|[\bi{\upphi}_{b,k}]_i|\bigg\}
\label{A3b}\\&=\displaystyle\max_{\|{\vec v}_{3b,3k}\|\leq
\rho_{b,k}^{\prime}} \sum_{i=1}^{l_{b,k}}\bigg\{\max\big(-f_2({\vec
M}_b,\boldsymbol\delta_{b,k}^i),-f_2({\vec
M}_b,-\boldsymbol\delta_{b,k}^i),0\big)|[{\vec v}_{3b,3k}]_i|\bigg\}\label{A3c}
\end{align}
\end{subequations}
where in \eqref{A3c} we have employed the result in optimization problem formulation \eqref{A2}, and have slightly
changed the representation in \eqref{21} by not specifying any particular norm in the constraint set. Now recalling the
definition of the dual norm $\vartheta\|\boldsymbol
\upgamma\|^\star\triangleq\displaystyle\max_{\|{\vec s}\|\leq
\vartheta}{\vec s}\trans\boldsymbol \upgamma$ of vector $\boldsymbol \upgamma$, we obtain the result stated in Theorem \ref{th1} for the
constraint set of interest.
\bibliographystyle{IEEEtran}
%\singlespacing
\bibliography{IEEEabrv,MISO_IC_RB}

\end{document}